\documentclass{pasj01}
\draft

\begin{document}
\SetRunningHead{}{}

\title{New Identification of the Mixed-Morphology 
Supernova Remnant G298.6$-$0.0 with Possible Gamma-ray Association}


\author{Aya \textsc{Bamba}$^1$,
Makoto \textsc{Sawada}$^1$,
Yuto \textsc{Nakano}$^1$,
Yukikatsu \textsc{Terada}$^2$,
John \textsc{Hewitt}$^{3,4}$,
Robert \textsc{Petre}$^3$,
Lorella \textsc{Angelini}$^3$,
}
\affil{$^1$
Department of Physics and Mathematics, Aoyama Gakuin University,
5-10-1 Fuchinobe Chuo-ku, Sagamihara, Kanagawa, 252-5258, Japan}
\affil{$^2$
Department of Physics, Science, Saitama University, Sakura, Saitama 338-8570,
Japan
}
\affil{$^3$
NASA Goddard Space Flight Center, Greenbelt, MD 20771, USA
}
\affil{$^4$
Center for Research and Exploration in Space Science and Technology (CRESST), University of Maryland, Baltimore County, Baltimore, MD 21250, USA
}


\maketitle

\begin{abstract}
We present an X-ray analysis on the Galactic supernova remnant
(SNR) G298.6$-$0.0 with Suzaku.
The X-ray image shows a center-filled structure inside the radio shell,
implying this SNR is categorized as a mixed-morphology (MM) SNR.
The spectrum is well reproduced by a single temperature plasma model
in ionization equilibrium,
with a temperature of 0.78 (0.70--0.87)~keV.
The total plasma mass of 30~M$_\odot$ indicates that
the plasma has interstellar medium origin.
The association with a GeV gamma-ray source 3FGL~J1214.0$-$6236 
on the shell of the SNR
is discussed,
in comparison with other MM SNRs with GeV gamma-ray associations.
It is found that the flux ratio
between absorption-corrected thermal X-rays and GeV gamma-rays decreases 
as the MM SNRs evolve to larger physical sizes.
The absorption-corrected X-ray flux of G298.6$-$0.0 and the GeV 
gamma-ray flux of 3FGL~J1214.0$-$6236
closely follow this trend, implying that 
3FGL~J1214.0$-$6236 is likely to be the GeV counterpart of
G298.6$-$0.0.
\end{abstract}


\section{Introduction}

The origin of Galactic cosmic rays has remained an unresolved question
since their discovery. One of the most plausible acceleration sites is
at the shock front of supernova remnants (SNRs).
The discovery of synchrotron X-rays \citep{koyama1995}
and very high energy (VHE) gamma-rays \citep{aharonian2004}
from shells of SNRs gave some of the first clear evidence of 
active particle acceleration in these shocks.
The small-scale structure and time variability of synchrotron X-rays 
imply that the acceleration is very efficient
\citep{bamba2003,bamba2005,uchiyama2007}.
However, it remains an open question how 
accelerated particles escape from the shocks to become cosmic rays.

Recently, Fermi detected GeV gamma-rays from several SNRs
\citep{abdo2010a,abdo2010b,abdo2010c}.
Interestingly, many of the GeV-detected SNRs are many times older 
than the non-thermal X-ray-detected or VHE-gamma-ray-detected SNRs
\citep{funk2011}.
Moreover, their GeV spectra are soft and show a spectral break at
energies around 1--10~GeV,
implying that high energy particles have already lost their energy
or have escaped from the acceleration sites.
If the latter is true, their soft spectra show the evidence of
particle escape. However, a larger sample of GeV gamma-ray 
SNRs is needed
to judge the particle escape scenario.

We searched for GeV gamma-ray sources
associated with cataloged Galactic SNRs in the 
Fermi Large Area Telescope Third Source Catalog \citep{3fgl}.
There are 60 GeV gamma-ray sources which spatially coincide 
with the Galactic SNRs, in addition to the 12 SNRs identified by 
extended gamma-ray emission.
3FGL~J1214.0$-$6236 is associated with the radio SNR G298.6$-$0.0,
 detected at 408~MHz and 843~MHz with a flat radio spectral index of $-$0.3
\citep{shaver1970,kesteven1987,whiteoak1996}.
\citet{reach2006} reported a possible detection of infrared emission
from the direction of G298.6$-$0.0, implying that the shock of this remnant
may have encountered a high-density medium. 
These characteristics are common among
GeV gamma-ray-emitting SNRs.
However, no detection has been made in the X-ray band
with ROSAT PSPC \citep{hwang1994}.
X-ray information is critical to estimate
the physical parameters of the SNR, such as 
age, explosion energy, and the amount of accelerated electrons.
Thus, our aim is to identify X-rays from G298.6$-$0.0 and measure parameters
using the low and stable background X-ray observations provided by Suzaku
\citep{mitsuda2007}.

In this paper, we report the first X-ray imaging spectroscopy of 
G298.6$-$0.0 using Suzaku.
The observation details are summarized in \S\ref{sec:obs}.
An analysis of the Suzaku X-ray data is
presented in \S\ref{sec:results},
the results of which are discussed in \S\ref{sec:discuss}.

\section{Observations and Data Reduction}
\label{sec:obs}

Suzaku observed G298.6$-$0.0 on 2012, Aug.\ 11--13
(sequence number: 507037010)
and 2013, Feb.\ 18--19 (sequence number: 507037020).
Only three set of the onboard X-ray Imaging Spectrometers (XIS0, 1, 3;
\cite{koyama2007}) 
could be used in this paper, due to a known problem with XIS2.
XIS1 is a back-illuminated CCD, whereas the others are front-illuminated.
The XIS instruments were operated
in the normal-clocking full-frame mode.
Spaced-row charge injection technique and the relevant energy calibration
method were applied to mitigate the long-term degradation of
the energy resolution
\citep{uchiyama2009}.
Since the observing mode is the same for the two observations,
we analyzed the combined data 
using HEADAS software version 6.16
and XSPEC version 12.8.2.
We reprocessed the data set with
the calibration database version 2012-11-06 for XIS
and 2011-06-30 for X-ray Telescope (XRT; \cite{serlemitsos2007}).

In the XIS data screening,
we filtered out data acquired during passages
through the South Atlantic Anomaly (SAA),
with elevation angles with respect to the Earth's dark limb below 5~deg,
or with elevation angle to the bright limb below 20~deg
in order to avoid contamination by emission from the bright limb.
The remaining exposure is 57~ks
(17~ks for the former observation and 40~ks for the latter observation).

\section{Results}
\label{sec:results}

\subsection{Images}

\begin{figure}
  \begin{center}
    \includegraphics[width=80mm]{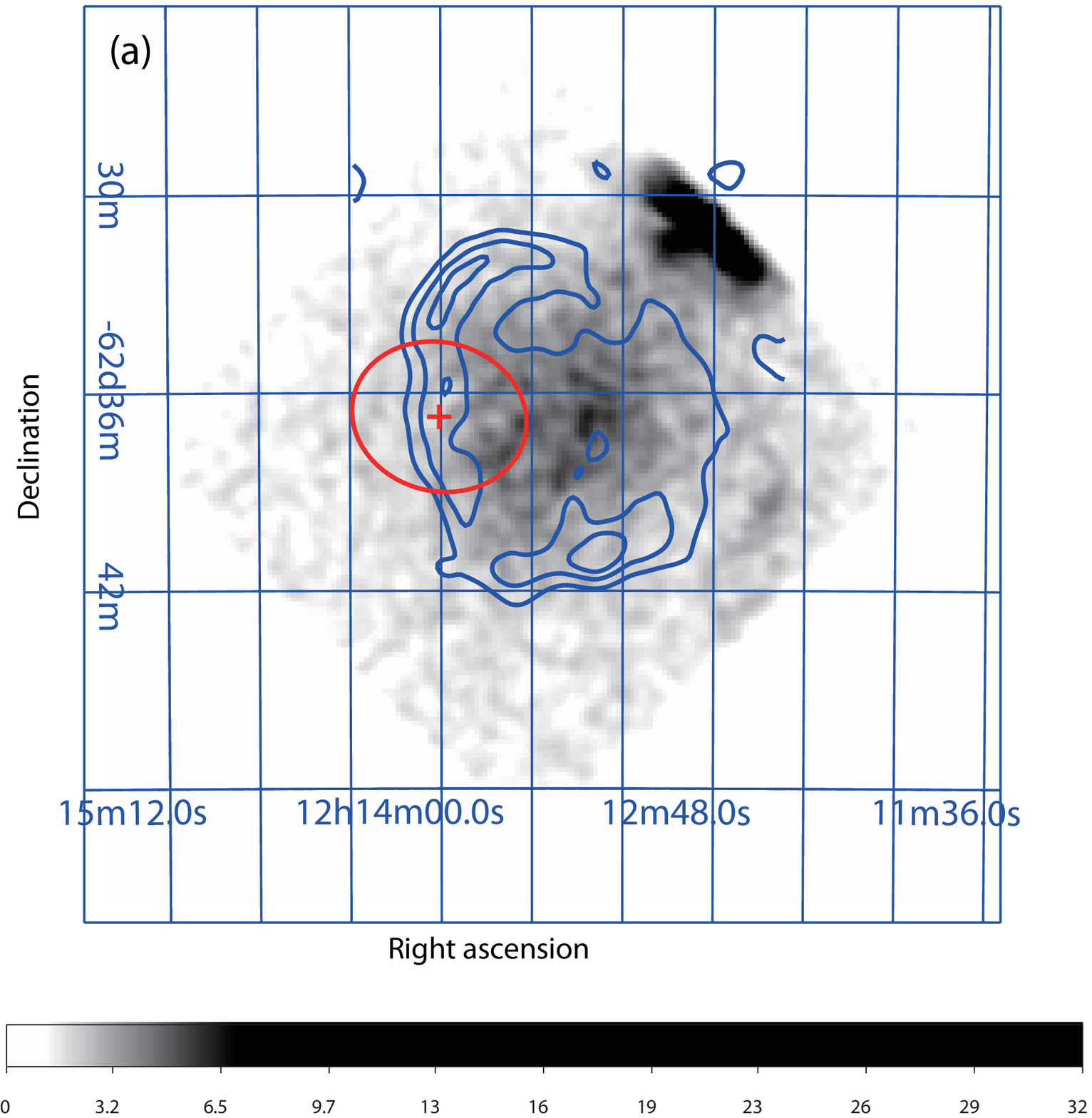}
    \includegraphics[width=80mm]{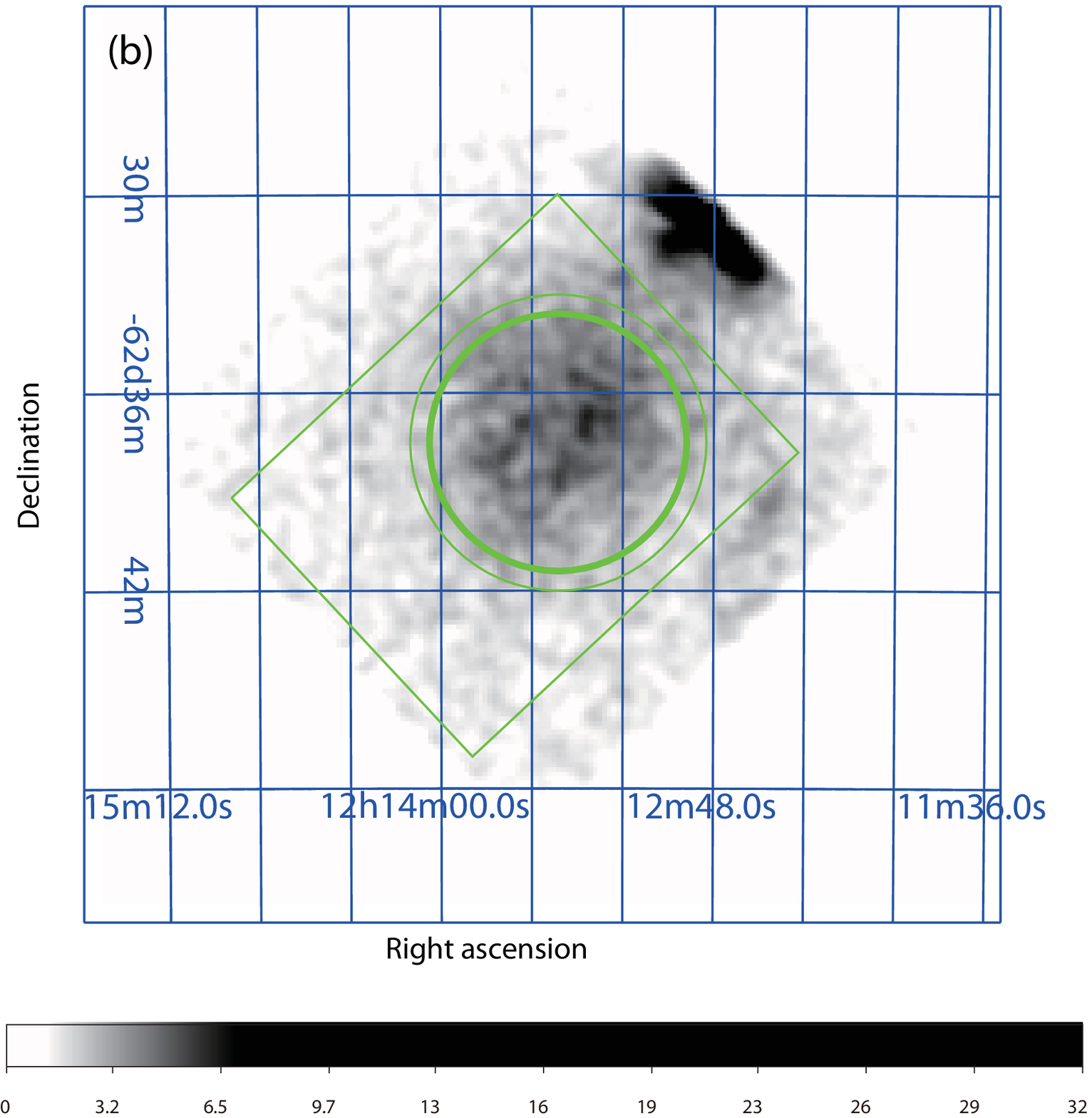}
  \end{center}
  \caption{(a): The 843~MHz (contours) and 0.5--5~keV (gray scale) images
of the G298.6$-$0.0 region.
The X-ray image is smoothed with a Gaussian function with the kernel of
0.4~arcmin,
Non--X-ray background (NXB) was not subtracted,
and no vignetting correction was performed.
The color scale is in logarithmic scale.
Red circle shows the error region of the Fermi source
3FGL~J1214.0$-$6236.
The coordinate is in J2000.
The point source in the north-west is 1RXS~J121248.7$-$623027
\citep{fuhrmeister2003}.
(b): Same image to (a) with source (bold) and background (thin) regions
for the spectral analysis.
}
  \label{fig:images}
\end{figure}

Figure~\ref{fig:images} (a) shows 843~MHz 
\citep{whiteoak1996} and the XIS~1+3 0.5--5~keV band
images of G298.6$-$0.0 region.
XIS~0 image is not used due to the dead column.
The position of the Fermi source, 3FGL~J1214.0$-$6236, is also shown.
The radio emission shows the clear shell-like structure,
while X-ray emission has a center-filled morphology
inside the radio shell and does not show any prominent point sources.
These results show that G298.6$-$0.0 is either 
a SNR with a central pulsar or a mixed-morphology SNR
(MM SNR; \cite{rho1998}).
The point source located in the north-east of the SNR is
coincident with the ROSAT source
1RXS~J121248.7$-$623027 \citep{fuhrmeister2003}
and categorized as a rotationally variable star \citep{kiraga2012},
whose details are out of the scope of this paper.

\subsection{Spectra}

The source spectra of G298.6$-$0.0 were extracted
from circular region with the radius of 3.9~arcmin,
whereas the background spectra were taken from
the source free region, as shown in Figure~\ref{fig:images} (b).
We added XIS0 and XIS3 spectra for the better statistics,
whereas XIS1 spectrum was treated separately
due to the different response
from the others.

\begin{figure}
  \begin{center}
    \includegraphics[width=80mm]{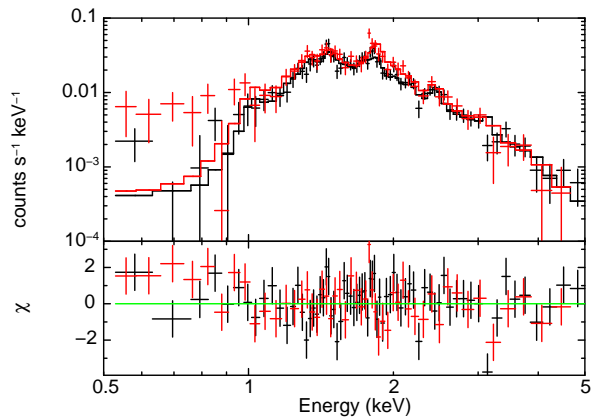}
  \end{center}
  \caption{%
Background-subtracted spectra of the source region.
Black and red closses represent FI and BI spectra, respectively.
The solid lines represent the best-fit {\sc vapec} model.
}
  \label{fig:spectra}
\end{figure}

Background-subtracted spectra are shown in Figure~\ref{fig:spectra}.
The spectra are rather soft and have emission lines,
implying that the emission has a thermal origin.
We thus adopted an optically thin thermal plasma model
in collisional ionization equilibrium ({\sc vapec})
affected by interstellar absorption
({\sc phabs}; \cite{balcinska-church1992}).
For both the emission and absorption models,
the solar abundance values of \citet{anders1989} were used.
The first fit with fixed abundances at the solar values
was rejected with the reduced $\chi^2$ of 228.6/117
and wavy residuals.
We thus fit with the abundances of Mg, Si, S, and Fe as free parameters.
This improved fit gave a reduced $\chi^2$ of 136.1/113.
The best-fit models and parameters are shown in 
Figure~\ref{fig:spectra} and Table~\ref{tab:spectra}, respectively.
We also tried to fit with non-equilibrium ionization plasma models,
{\sc vnei} for ionizing plasma and {\sc vvrnei} for recombining plasma.
The abundances of Mg, Si, S, and Fe were treated as free parameters.
For a {\sc vvrnei} model, we fixed the initial plasma temperature to be
3~keV.
The fitting showed similar reduced $\chi^2$ of
136.5/112 ({\sc vnei}) and 131.5/112 ({\sc vvrnei}).
The best-fit values of the ionization/recombination timescale are
$>9.8\times 10^{11}$~cm$^{-3}$s
for the {\sc vnei} model
and 6.2 (5.3--14) $\times 10^{11}$~cm$^{-3}$s
for the {\sc vvrnei} model, respectively.
The latter gives a slightly shorter relaxation timescale
than that in ionization equilibrium ($10^{12-13}$~s~cm$^{-3}$).
However, the improvement of the fit is only marginal;
both models show similar reduced $\chi^2$.
Moreover, there is no clear indication of strong radiative recombination
continuum in the residual
which is the direct evidence of a recombining plasma \citep{yamaguchi2009}.
Thus 
we concluded that the plasma is well described by 
an absorbed plasma emission model in ionization equilibrium.
We note that the XIS1 spectrum has residuals in the low energy band,
which may be due to the calibration uncertainty of contamination.

\section{Discussion}
\label{sec:discuss}

\subsection{Physical Parameters of G298.6$-$0.0}

We report the first detection of thermal X-rays from inside the radio
SNR G298.6$-$0.0. The characteristics of G298.6$-$0.0 
classify it as a new MM SNR.
Here, we estimate several physical parameters for this SNR.

First, we estimate the distance to this source.
We measured the absorption column to the source to be
$1.65~(1.45-1.99)\times 10^{22}$~cm$^{-2}$.
The total hydrogen density throughout the Galaxy
in this direction is estimated by 
the method of \citet{willingale2013}.
The ${\rm H}_{\rm I}$ column density in this direction
is 1.4~$\times 10^{22}$~atoms~cm$^{-2}$ \citep{kalberla2005},
whereas the column density of molecular hydrogen (${\rm H_2}$) is
1.4$\times 10^{21}$~atoms~cm$^{-2}$ from the dust map \citep{schlegel1998}.
The total absorption column is thus
1.5$\times 10^{22}$~atoms~cm$^{-2}$, which is similar value to our target.
This indicates that the SNR is located in the outer part of the Galaxy.
According to the face-on map by \citet{nakanishi2006},
most of the interstellar hydrogens in this direction are located within 10~kpc
from the Sun.
Thus we adopt the distance of 10~kpc.
This is an independent constraint from that indicated
by the $\Sigma-D$ relation in the radio band (9.5~kpc; \cite{case1998}).
The previous method only gives
a rough estimate and more detailed
studies are necessary to get a better measurement of
the distance to SNR G298.6$-$0.0.

Assuming the plasma fills a 3.9~arcmin sphere,
the total volume is 1.8$\times 10^{59}D_{10}^3$~cm$^{3}$,
where $D_{10}$ is the distance in the unit of 10~kpc.
The average density derived from the emission measure
(Table~\ref{tab:spectra})
is 0.2$D_{10}^{-1/2}f^{1/2}$~cm$^{-3}$, where $f$ is the filling factor.
The total mass and thermal energy can be estimated as
30$D_{10}^{5/2}f^{1/2}M_\odot$ and $6.4\times 10^{49}D_{10}^{5/2}f^{1/2}$~erg,
respectively. The large total mass indicates that
the main component of the plasma originates from 
swept-up interstellar medium.

\subsection{Comparison with other SNRs with GeV counterpart}

Here, we discuss on the comparison with G298.6$-$0.0 and
other MM SNRs with GeV gamma rays,
to judge
whether 3FGL~J1214$-$6236 has similar properties
to other GeV counterparts of MM SNRs.

GeV emission is coincident with the shell of G298.6$-$0.0,
which is same to other SNR cases \citep{abdo2010a}.
The possible interaction with a high-density medium
from the direction of this SNR is reported \citep{reach2006},
which is also similar to other GeV SNR samples.
On the other hand, we have no information 
which part of this SNR interacts with high-density medium.
Further studies in infrared or molecular cloud bands is needed
to conclude the interaction.
The GeV spectrum of 3FGL~J1214.0$-$6236 has a spectral break
around a few GeV \citep{3fgl},
which is also similar to the spectra of other GeV SNRs.
These facts imply that 3FGL~J1214.0$-$6236 is a possible GeV counterpart
of G298.6$-$0.0.

Table~\ref{tab:comparison} shows the summary of the 
properties across the
absorption-corrected 0.3--10~keV and 0.1--100~GeV bands of the 
GeV gamma-ray emitting MM SNRs.
Large distance uncertainties for these sources make
it difficult to apply a detailed model of the X-ray and GeV gamma-ray 
evolution to the case of G298.6$-$0.0. Instead we compare the 
flux ratio between absorption-corrected X-rays and GeV gamma-rays,
which is independent from the distance.

We selected samples with known
total absorption-corrected X-ray and GeV gamma-ray fluxes.
Recent studies revealed that
some of our samples have X-ray emission
from recombining plasma with detailed analysis
\citep{ozawa2009,uchida2012,sato2014},
but we concentrate on total X-ray flux from older results
as it is an easy comparison for the other fainter sources.
IC443 is a famous MM SNRs with GeV gamma-rays
\citep{yamaguchi2009,abdo2010c},
but do not have a good estimate of the total X-ray flux in the literatures
due to their large apparent diameters,
and thus we did not adopt it as our samples.
We also ignored emission from ejecta reported in W49B and G349.7+0.2
\citep{kawasaki2005,yasumi2014},
which contributes only $\sim$10\% flux of the total flux.
GeV gamma-ray spectra for several sources have spectral breaks
around 10~GeV
\citep[for example]{abdo2010a,abdo2010b,abdo2010c}.
To treat fainter sources equally,
we used the total energy flux from 
the Fermi source catalog \citep{3fgl}.
The difference in the fluxes of W44 between the catalog and that derived by
\citet{abdo2010a}
is only about 20\%,
which is negligible for our study.
We estimated the radii using the catalog of 
Galactic SNRs by \citet{green2014}.

\begin{figure}
  \begin{center}
    \includegraphics[width=80mm]{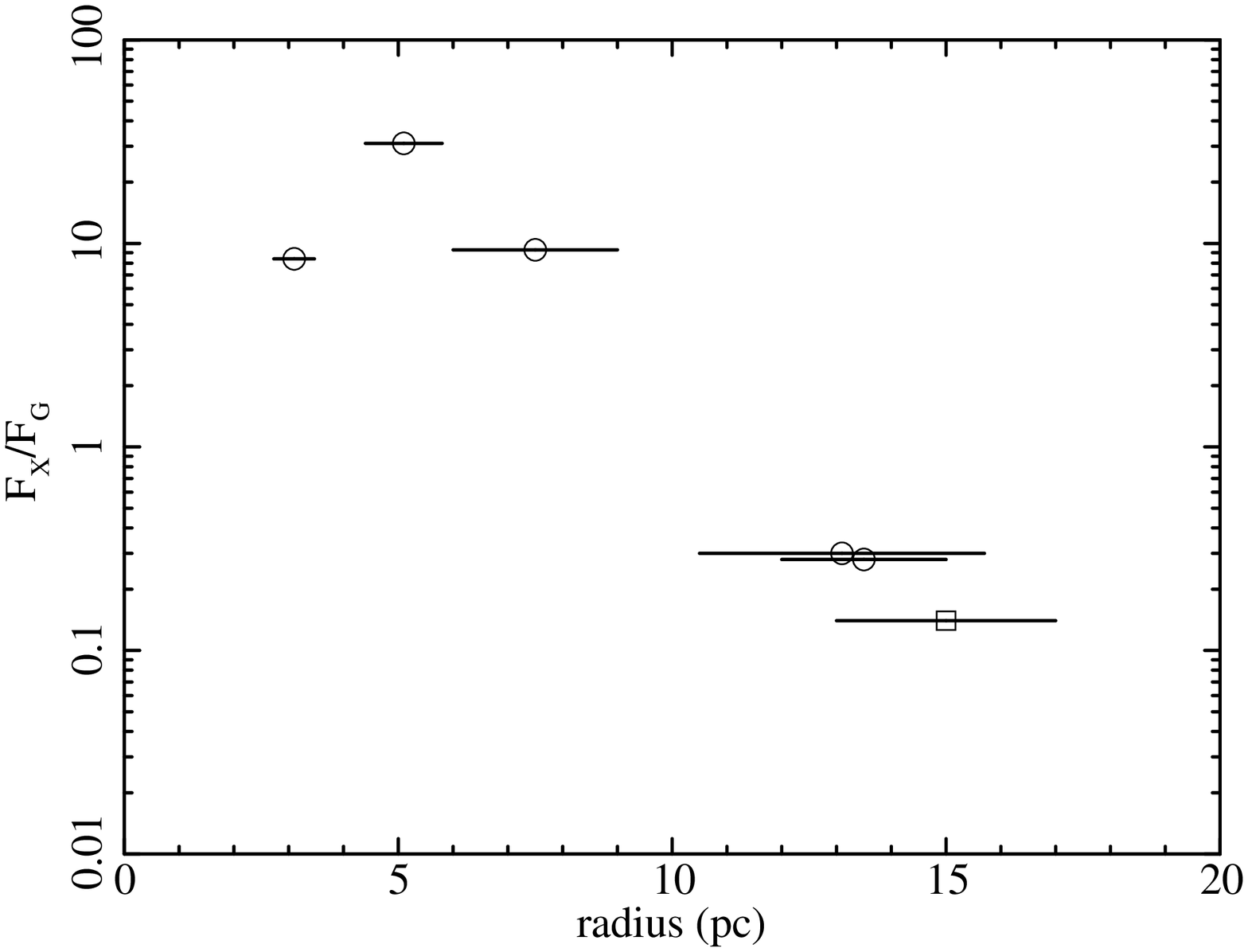}
    \includegraphics[width=80mm]{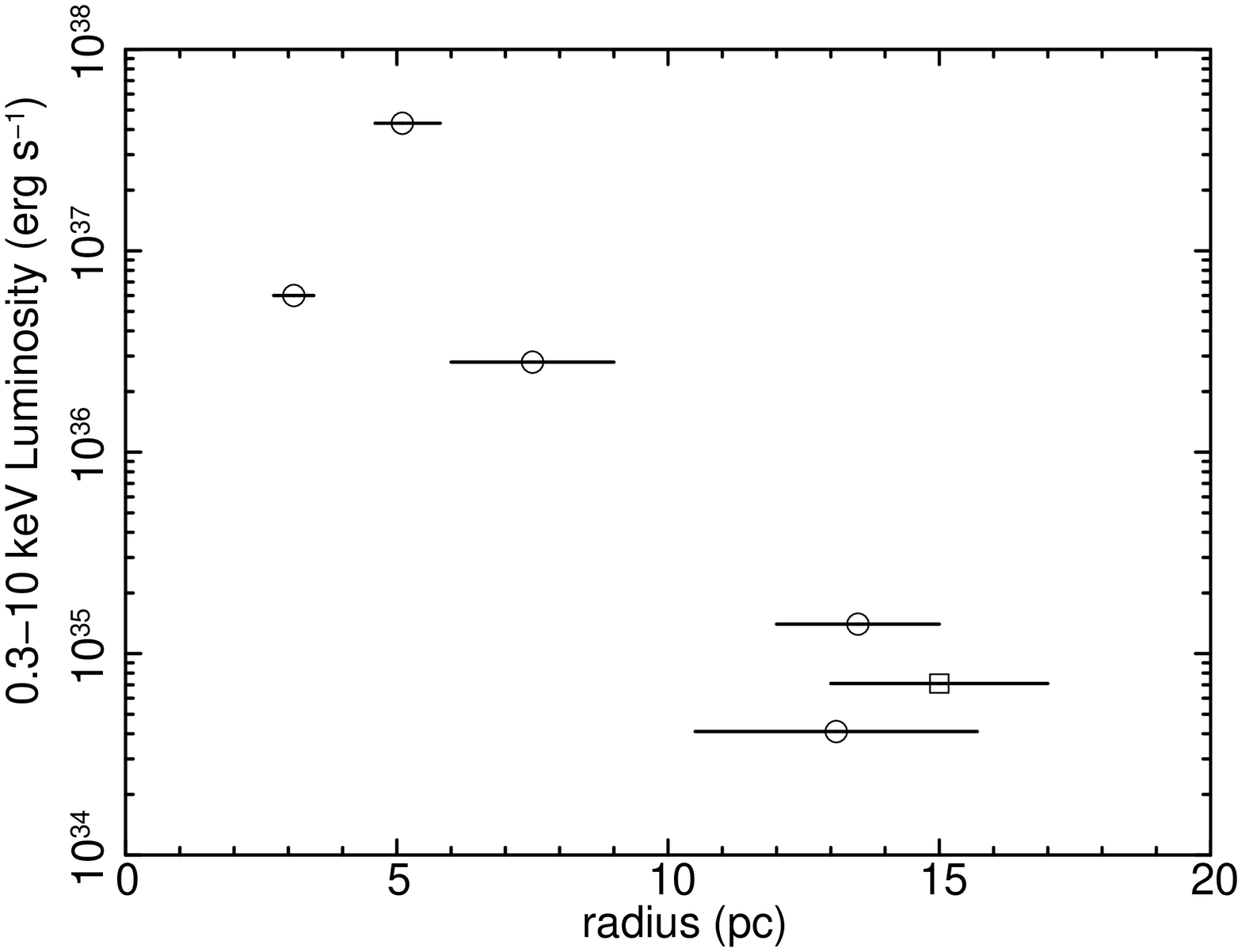}
    \includegraphics[width=80mm]{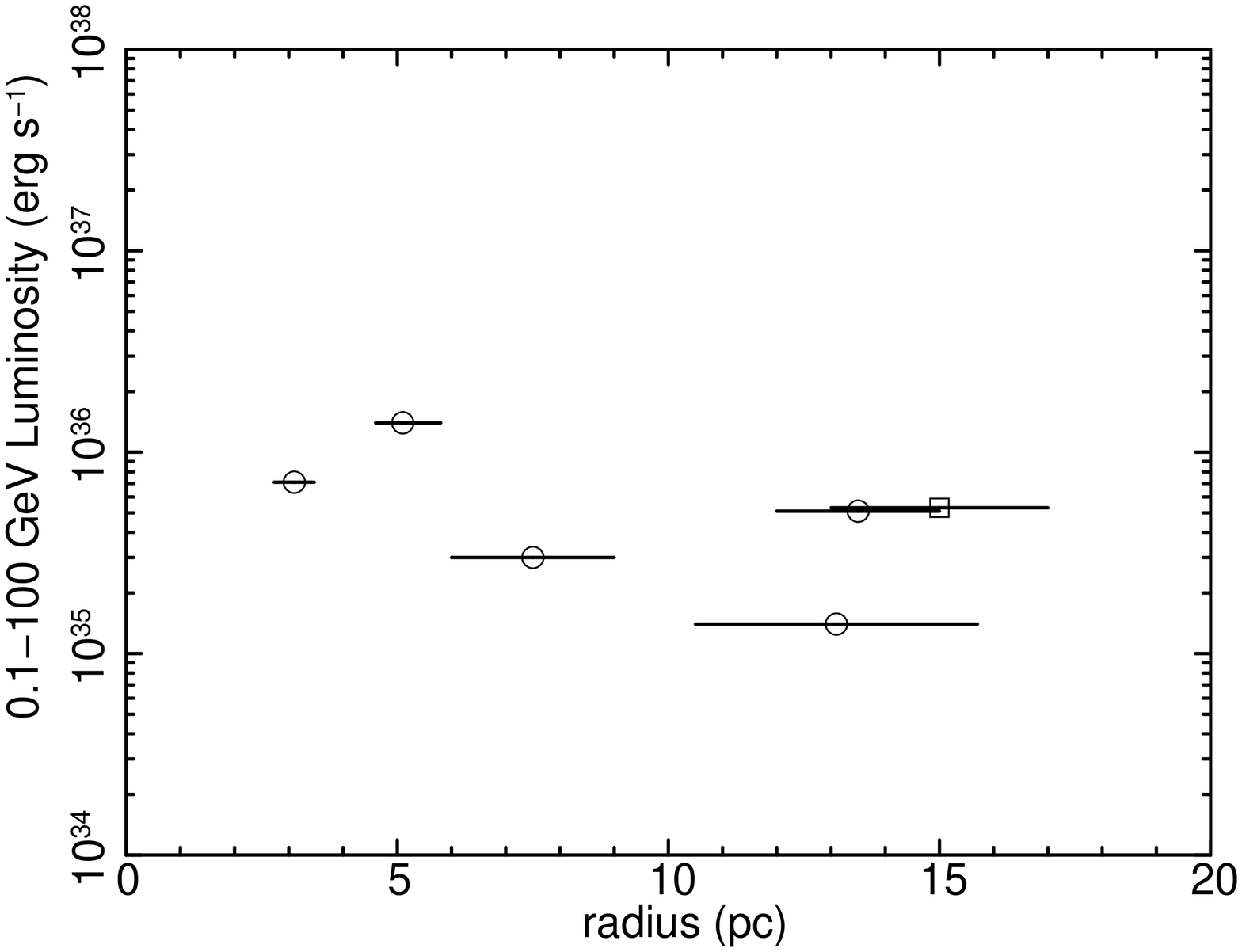}
  \end{center}
  \caption{%
(a):
Radius vs. flux ratio between 0.3--10~keV (absorption-corrected, $F_X$)
and 0.1--100~GeV ($F_G$)
for MM SNRs with GeV gamma-ray emission.
Circles and the box represent the samples from literatures
and from this work, respectively, in all panels.
(b): Radius vs. 0.3--10~keV luminosity.
(c): Radius vs. 0.1--100~GeV luminosity.
}
  \label{fig:r-ratio}
\end{figure}

The flux ratio between 0.3--10~keV and 0.1--100~GeV,
$F_X/F_G$ in Table~\ref{tab:comparison},
is in the range of $\sim$0.1--30.
If 3FGL~J1214.0$-$6236 is the gamma-ray counterpart of G298.6$-$0.0,
our target has the faintest $F_X/F_G$ ratio.
Figure~\ref{fig:r-ratio}(a) shows the
radius vs. $F_X/F_G$ of MM SNRs with GeV gamma-rays.
The flux ratio shows decline as the radii of SNRs is larger,
or SNRs evolves older.
The flux ratio of our target is well consistent with this trend.

It is unclear why the flux ratio becomes smaller when MM SNRs evolve.
One possibility is that the X-ray luminosity decreases faster
due to the expansion and cooling
\citep{gehrels1993},
whereas the GeV gamma-ray luminosity does not decrease.
In order to check this hypothesis,
we have made plots of radius vs. 0.3--10~keV luminosity
(Figure~\ref{fig:r-ratio}(b))
or 0.1--100~GeV luminosity (Figure~\ref{fig:r-ratio}(c)).
This is consistent with our scenario;
The X-ray luminosity become smaller when SNRs evolve (or radii become larger),
whereas the gamma-ray luminosity is almost constant in the range of 
$10^{35}$--$10^{36}$~erg~s$^{-1}$.
This is also consistent with previous results
in X-rays \citep{long1983} and GeV gamma-rays \citep{brandt2015}.
The constant gamma-ray flux support our hypothesis
that timescale for low-energy particles to escape
these SNRs is long in comparison to the SNR evolution.
This trend also implies that
a part of Galactic GeV unID sources can be very evolved MM SNRs
without significant detection in X-ray and radio bands.
Note that radio surface brightness becomes smaller when the SNR evolves
\citep{case1998},
which makes difficult to detect them together with large angular size.
Further study of a large sample of GeV gamma-ray SNRs
is needed to understand this tendancy.

\begin{ack}
We thank the anonymous referee for the fruitful comments.
This research has made use of the SIMBAD
database, operated at CDS, Strasbourg, France.
This work was supported in part by
Grant-in-Aid for Scientific Research
of the Japanese Ministry of Education, Culture, Sports, Science
and Technology (MEXT) of Japan, No.~22684012 and 15K05107 (A.~B.)
and No.~15K17657 (M.~S.).
\end{ack}

\begin{table}
\caption{Best-fit parameters of the spectral fitting ({\sc vapec})$^\dagger$.}
\label{tab:spectra}
\begin{center}
\begin{tabular}{p{8pc}c}
\hline\hline
Parameters & Values \\ \hline
$N_{\rm H}$ ($10^{22}$~cm$^{-2}$)\dotfill & 1.65 (1.45--1.99) \\
$kT$ (keV)\dotfill & 0.78 (0.70--0.87) \\
Mg\dotfill & 0.59 (0.39--0.86) \\
Si\dotfill & 0.38 (0.27--0.50) \\
S\dotfill & 0.35 (0.21--0.50) \\
Fe\dotfill & $<$0.23 \\
$E.M.^\ddagger$\dotfill & 5.6 (4.1--8.1) \\
reduced $\chi^2$\dotfill & 136.1/113 \\
\hline
\multicolumn{2}{p{20pc}}{$^\dagger$: Errors indicate single parameter 90\% confidence interval.}\\
\multicolumn{2}{p{20pc}}{$^\ddagger$: Emission measure in the unit of
$\frac{10^{-11}}{4\pi D^2}\int n_en_HdV {\rm cm}^{-5}$, where
$D$, $n_e$, and $n_H$ represent the distance,
and electron and hydrogen densities.}
\end{tabular}
\end{center}
\end{table}

\begin{table}
\caption{X-ray and gamma-ray fluxes of MM SNRs with associations to GeV gamma-rays}
\label{tab:comparison}
\begin{center}
\begin{tabular}{p{4.5pc}cccccc}
\hline\hline
Sample & $D^a$ & Radius$^b$ & $F_X{}^c$ & $F_G{}^d$ & 
$F_X/F_G$ & References \\
\hline
W 49B\dotfill & 8 & $5.1\pm0.7$ & 56$^e$ & 1.8 & 31 & (1) (2) (3) \\
G349.7+0.2\dotfill & 11 & $3.0\pm0.4$ & 4.1$^e$ & 0.49 & 8.4 & (4) (5) (3) \\
3C 391\dotfill & 7.2 & $7.5\pm1.5$ & 4.5 & 0.49 & 9.3 & (6) (2) (3) \\
W28\dotfill & 1.8 & $13.1\pm2.6$ & 1.1$^f$ & 3.6 & 0.30 & (7) (8) (3) \\
W 44\dotfill & 2.8 & $13.5\pm1.5$ & 1.5$^f$ & 5.4 & 0.28 & (1) (2) (3) \\
G298.6+0.0\dotfill & 10 & $15\pm2$ & 0.059 & 0.44 & 0.14 & this work, (3) \\
\hline
\multicolumn{7}{p{25pc}}{Note --- 
(1) \citet{claussen1997},
(2) \citet{kawasaki2005}, 
(3) \citet{3fgl},
(4) \citet{tian2014},
(5) \citet{yasumi2014},
(6) \citet{radhakrishnan1972},
(7) \citet{clemens1985},
(8) \citet{rho2002}
}\\
\multicolumn{7}{p{25pc}}{$^a$:Distance to the source in the unit of kpc.}\\
\multicolumn{7}{p{25pc}}{$^b$: In units of pc. Errors indicate the radii along the major and minor axes.}\\
\multicolumn{7}{p{25pc}}{$^c$: absorption-corrected 0.3--10~keV flux in units of $10^{-10}$~erg~cm$^{-2}$s$^{-1}$.}\\
\multicolumn{7}{p{25pc}}{$^d$: 0.1--100~GeV flux in units of $10^{-10}$~erg~cm$^{-2}$s$^{-1}$.}\\
\multicolumn{7}{p{25pc}}{$^e$: Only ISM component was considered.}\\
\multicolumn{7}{p{25pc}}{$^f$: The X-ray flux was not extracted
from the entire SNR but from the brightest central regions.}\\
\end{tabular}
\end{center}
\end{table}

\end{document}